\documentclass[aps,preprint]{revtex4}
\usepackage{graphicx}

\begin{document}

\title{Wide viewing-angle holographic display based on enhanced-NA Fresnel hologram}

\author{Byung Gyu Chae}

\address{Holographic Contents Research Laboratory, Electronics and Telecommunications Research Institute, 218 Gajeong-ro, Yuseong-gu, 
Daejeon 34129, Republic of Korea}

%\date{\today}

\begin{abstract}

The viewing-angle enlargement of a holographic image is a crucial factor for realizing the holographic display.
The numerical aperture (NA) of digital hologram other than a pixel specification has been known to determine the angular field extent of image.
Here, we provide a valid foundation for the dependence of viewing angle on the hologram numerical aperture
by investigating mathematically the internal structure of the sampled point spread function showing a self-similarity of its modulating curves.
The enhanced-NA Fresnel hologram reconstructs the images at a viewing angle larger than a diffraction angle by a hologram pixel pitch
where its angle value is expressed in terms of the NA of whole hologram aperture,
which is systematically observed by optical hologram imaging.
Finally, we found that the aliased replica noises generated in the enhanced-NA Fresnel diffraction regime are effectively suppressed within the diffraction scope by a digitized pixel.
This characteristic enables us to overcome the image reduction and to remove the interference of high-order images, which leads to the wide viewing-angle holographic display.
Optical experiments are shown to be consistent with the results of numerical simulation.

\end{abstract}
%\pacs{42.30.Kq, 42.40.-i, 42.25.Fx, diffraction optical 05.45.Df, fractals}
\maketitle{}

\section{Introduction}
The holographic display is an optical imaging system that reconstructs the real or imaginary image in a free space from the digital hologram \cite{1,2,3,4}.
The spatial resolution of the numerically reconstructed image in digital holography is subject to the Abbe-Rayleigh diffraction limit related to the hologram numerical aperture \cite{5,6,7}.
We studied the angular field view of optically reconstructed image in holographic display by investigating the diffraction fringes propagating
from the sampled hologram loaded on a pixelated spatial light modulator \cite{8}.
We found that the viewing angle of holographic image is fundamentally determined by the numerical aperture (NA) of digital hologram,
whereas the pixelated structure invokes only the generation of high-order images.
That is, the viewing angle changes with an image position even in the spatial light modulator with a finite pixel pitch.
The viewing angle $\it{\Omega}$ depends on only the geometrical structures of lateral size, $L = N\it{\Delta} x$ and distance $z$:
\begin{equation}
\it{\Omega} = \rm{}2 \sin^{-1} \left( \frac{\it{N} \it{\Delta} \it{x}}{2\it{z}} \right).
\end{equation}
The angle value does not directly relate to the diffraction ability from the pixel size $\it{\Delta} x$ or a wavelength $\lambda$ of incident wave.
This appears to be unintuitive in some ways because the lateral scope of diffractive wave from the pixel would define the viewing zone.
However, we investigated in detail that the variation of viewing angle well obeys the relation of Eq. (1) \cite{8}.

The extension of viewing angle is a crucial task for realizing the holographic display \cite{9,10,11}.
Based on Eq. (1), the hologram numerical aperture can be enhanced in the digital hologram synthesized at a closer distance.
When the digital hologram is made at a distance lower than a critical distance, $z_c = N \it{\Delta} x^{\rm{2}} /\lambda$,
the viewing angle would be higher than the diffraction angle by a pixel pitch.
We defined this type of hologram as the enhanced-NA Fresnel hologram \cite{12}.
In such circumstance, the hologram fringe undergoes the aliasing errors when being sampled.
The sampling condition has been well understood in the Nyquist sampling theorem with respect to both planes or by the Wigner domain description \cite{13,14,15,16}.
The window extent of both planes is calculated by the pixel pitch of opposite plane as with a feature of simple Fourier transform,
where only a limited size object could be recoverable.
Onural and Stern et al. \cite{17,18} researched that from the Fourier analysis of sampling theory,
the object smaller than the diffraction range by a hologram pixel can be fully reconstructed even in severe aliased environment.

We also studied the aliased error of hologram fringe and the image recovery from this type of hologram \cite{12}.
The aliased fringes are due to the undersampling of high-frequency components, which appears in the form of replica patterns.
Here, the object size decreases even though the viewing angle increases.
When the object larger than the diffraction extent by a hologram pixel pitch is used, the interference of high-order images is inevitable.
Particularly, we found that the replica fringes in the enhanced-NA hologram are nearly suppressed due to the near-field diffraction property,
which gives an opportunity for accomplishing a viewing-angle enlargement without contracting the image size.

Generally, the resolution of the numerically reconstructed image has been interpreted by the window function covering the whole hologram aperture.
Despite some related researches, there remains a question about defining the hologram numerical aperture in optical reconstruction.
The deficiency of high-frequency components may affect the image formation.
Therefore, a further study is required to clarify the definition of NA in above type of digital hologram.

In this study, we investigate mathematically the internal structure of the sampled point spread function (PSF) showing a self-similarity of its modulating curve.
The PSF becomes a single Fresnel zone forming the object image.
The diffraction-limited resolution relates to the resolving ability of two closest point images.
From this, we characterize the dependence of viewing angle on the hologram numerical aperture, especially based on quantum mechanical framework.
Subsequently, we carry out the optical experiments for systematically observing the viewing angle variation dependent on the hologram numerical aperture,
and finally, suggest a scheme for realizing a wide viewing-angle holographic display

\section{Viewing angle of holographic image dependent on hologram numerical aperture}
\subsection{Fractal structure of modulating curves in sampled point spread function}

The Fresnel diffraction field is well described by the convolutional kernel function $h(x,y)$ of the PSF \cite{1},
\begin{equation}
h(x,y) = \frac{e^{i kz}}{i\lambda z} \exp \left[i\frac{\pi}{\lambda z} (x^2 + y^2) \right],
\end{equation}
where $k$ is a wavenumber expressed as  $2\pi/\lambda$, and $z$ is a propagation distance.
The PSF reveals the interesting mathematical properties because of a quadratic phase term when being sampled.
The Fourier transform $\textbf{\textit{FT}}$ of sampled PSF in one-dimensional description is written by
\begin{equation}
\textbf{\textit{FT}} \left( \sum_{n} h(n\it{\Delta} x) \delta(x-n\it{\Delta} x) \right) = \frac{1}{\it{\Delta} x} \sum_{q} H \left( f - \frac{q}{\it{\Delta} x} \right),
\end{equation}
where $\it{\Delta} x$ is the sampling period. $H(f)$ is also a quadratic phase function of spatial frequency $f$,
which is called a optical transfer function.
The term in the summation of Eq. (3) can be expressed as the modulation of $H$ function, $c_{q/\it{\Delta} x}H(f)\exp (i 2\pi \lambda zqf / \it{\Delta} x)$
where $c_{q/\it{\Delta} x} = \exp (-i \pi \lambda zq^2 / \it{\Delta} x^{\rm{2}})$. From this, we obtained following equality \cite{19},
\begin{equation}
\sum_{n} h(n\it{\Delta} x) \delta(x-n\it{\Delta} x) = \frac{\rm{1}}{\it{\Delta} x} \sum_{q} c_{q/\it{\Delta} x} h \left( x + \frac{\lambda zq}{\it{\Delta} x} \right).
\end{equation}
The sampled PSF is represented by sum of continuous quadratic functions, 
which induces the replication of weighted original function with a period of $\lambda z/ \it{\Delta} x$,
where it does not make a meaningless aliased fringe when being sampled.
This implies that the PSF undersampled by $s$ multiples of $\it{\Delta} x$ forms the replica functions at a reduced period of $\lambda z/ (s\it{\Delta} x)$.
We can also extract the minimum distance $z_c$ for the function with $N$ samples without inducing replication.
When the interval of replica fringes is the same as a field extent, $N \it{\Delta} x$, $z_c = N \it{\Delta} x^{\rm{2}} /\lambda$.
In a distance greater than $z_c$, no aliased fringe appears within the field extent.

Figure 1 displays a quadratic sinusoid of sampled kernel function.
The magnitude of curves seems to be very irregular unlike the analog signal, but the curves include their internal structure.
If one divides the function of Eq. (4) into two parts, sampled by even numbers $n_e$ and odd numbers $n_o$, i.e.,
$\sum_{n_e} h(n_e\it{\Delta} x) \delta(x-n_e\it{\Delta} x) + \sum_{n_o} h(n_o\it{\Delta} x) \delta(x-n_o\it{\Delta} x)$,
each term also consists of replica functions with an interval $\lambda z/ (2\it{\Delta} x)$:
\begin{equation}
\left[c_0 h(x) + c_1 h \left(x + \frac{\lambda z}{2\it{\Delta} x} \right) + \cdots \right]_{even} + \left[c_0 h(x) + c_1 h \left(x + \frac{\lambda z}{2\it{\Delta} x} \right) + \cdots \right]_{odd}.
\end{equation}
Similarly, each term of Eq. (5) can be resampled into two parts, 
and when the function is successively sampled by an $s$-fold sampling period, the replica patterns divided by the $s$-fold are generated,
as depicted in Figs. 1(a) and 1(b).

Furthermore, the sampled kernel function has the modulating curves of original function according to its geometrical placement.
Let us incorporate two terms of Eq. (5).
The incorporation of two terms results in a twofold increase in sampling rate.
The primary Fresnelet centered on the axis simply becomes original form.
Meanwhile, the secondary Fresnelet should be made using the specifications ($z/2$, $\it{\Delta} x$) in a view of its placement \cite{12},
and thus, they would be modified in the form using a sampling rate $\it{\Delta} x^{\rm{-1}}$:
\begin{equation}
%\exp \left (-\frac{i\pi}{\lambda z} x^2 \right) \exp \left (-\frac{i\pi}{\Delta x} x \right) 
\exp \left[\frac{i\pi}{\lambda z/2} \left (x + \frac{\lambda z/2}{\it{\Delta} x}\right)^2 \right].
\end{equation}
The secondary Fresnelet has a phase coefficient of $(\lambda z/2)^{-1}$,
and appears as the modulation of complex exponential function,
which is clearly confirmed in the subsidiary curve of Figs. 1(a) and 1(b).
Since the width $\delta X$ of core shell of primary Fresnelet is given by
\begin{equation}
-\sqrt{\lambda z} \leq \delta X \leq \sqrt{\lambda z},
\end{equation}
the secondary Fresnelet has the core shell width of $(1/\sqrt{2})\delta X$, which is also confirmed in two-dimensional Fresnel zones of Fig. 2.
Likewise, the value in the subsidiary Fresnelets digitized by $s$-fold sampling period would be
\begin{equation}
(\frac{1}{\sqrt{s}})\delta X.
\end{equation}

Above characteristics induces a self-similar structure of modulating curves of PSF.
The similar subsidiary curves are recursively created in arbitrary sampled PSF,
depending on two internal variables ($z$, $\it{\Delta} x$) of distance and sampling period.
Figure 2 shows the fractal structure of two-dimensional Fresnelet for angle-valued kernel function.
The similar zones in a small scale are emerged when magnifying the pattern, where the number of smaller zones increases fourfold at half the scale parameter. 
This is a typical fractal characteristic with a Hausdorff fractal dimension of two \cite{20}.
However, we can observe that the structure in two-dimensional space is somewhat complicate.
Since the description based on Eq. (5) relates to only the multiple of sampling period $2\it{\Delta} x$,
other sampling periods as like $3\it{\Delta} x$ and $5\it{\Delta} x$ should be considered to analyze the similar patterns completely.
The subsidiary zones created by sampling period of prime number pixel $3\it{\Delta} x$ appear in Fig. 2,
which could also create their self-similar zones, where a Hausdorff dimension is still two.
We find that in the Fresnelet fractal, the fractal structures with respect to various sampling periods are mixed.

\subsection{Viewing angle analysis of reconstructed image from enhanced-NA Fresnel hologram based on quantum mechanical approach}

We investigate the property of viewing angle of the holographic image reconstructed from the enhanced-NA Fresnel hologram.
The object field $o(x',y')$ reconstructed from the digital hologram $g(x, y)$ with a finite lateral size $L$ is represented by,
\begin{equation}
o(x',y') = g(x',y') \rm{rect} \it{} \left(\frac{x'}{L},\frac{y'}{L}\right) \ast h^{\rm{-1}}(x',y').
\end{equation}
where $\ast$ indicates a convolutional operation and rect() is a rectangular function.
The definition of hologram aperture size should be cautiously considered because the enhanced-NA hologram forms the replica fringes.
The replica fringes can be analyzed by direct observation of the sampled hologram pattern for a point object that composes a real object \cite{12}.

In an on-axis point object of delta function,
the real or imaginary value of the kernel function $h(x,y)$ becomes a hologram in the form of a Fresnel zone.
Here, the replica patterns of Fresnel zone can be generated at below the $z_c$-distance.
In digital hologram synthesized at a half of $z_c$,
the modulating curve of Eq. (6) becomes in the form of Fresnel zones.
The fractal property of modulating curve makes adaptively the well-behaved replications of Fresnel zones in accordance with a decrease of distance.
The digital hologram for the object as a collection of points is synthesized by summing up the Fresnel zones \cite{21}.

Let us unwrap the phase hologram of point object.
As illustrated in Fig. 3(a), the unwrapped phase synthesized at a critical distance $z_c$ reveals the single Fresnel lens with smooth curvature.
The single Fresnelet acts like Fourier lens to focus the incident wave to a point spot.
In case of the enhanced-NA hologram synthesized at a distance lower than $z_c$, the multiple Fresnel lens are formed in Fig. 3(b). 
They can generate the multiple point images from the incident plane wave.

The minimum resolvable distance between two point objects is interpreted within the Abbe-Rayleigh diffraction theory.
The point image is reconstructed from only the Fourier transform of aperture window because the hologram having an optical kernel function is cancelled out by its inverse term.
We described the spatial resolution $R$ by hologram numerical aperture \cite{8}:
\begin{equation}
R = \frac{\lambda}{\textrm{2NA}}.
\end{equation}
Numerical aperture is expressed as the geometric structure of aperture size, $L=N\it{\Delta} x$ and distance $z$, $ \rm{NA} = \sin \it{\Omega}_{\rm{NA}} = N \Delta x /\rm{2}\it{z}$.
The total spatial extent $L$ of digital hologram becomes the aperture window in a single Fresnel lens,
whereas in case of multiple lens, it is not straightforward for defining the aperture window.
We know that even in the digital hologram properly sampled at critical distance $z_c$,
the symmetrically distributed point sources can generate the similar Fresnel patterns.
Here, it is natural that the total area plays a role in an aperture,
because the individual Fresnel patterns are spread over the whole area and overlapped each other.
However, as depicted in Fig. 1, the formation of multiple zones is due to the undersampling of high-frequency components in the Fresnelet.
This process is rather similar to that of high-order diffraction patterns by a relatively large pixel pitch.
Meanwhile, the components of each Fresnelet should be still the high-frequency components of adjacent Fresnelets.
The higher angular spectrum components of digital hologram survive.

We try a quantum mechanical approach to clarify the range of aperture window in the enhanced-NA hologram.
The photon field incident on the hologram aperture forms the multiple point images in the image plane.
Let us represent the spatial image mode by a density matrix $\rho$,
\begin{equation}
\rho = \sum_{q} e_q \mid \psi_q > <\psi_q \mid,
\end{equation}
where $e_q$ is probability.
This coefficient has the same quantity, because it has been studied that the point images have an equally distributed intensity \cite{22,23}.
The spatial mode state $\mid \psi>$  is given by a sinc or jinc function in accordance with an aperture shape. 
This state is calculated by the inverse transform of Eq. (9):
\begin{equation}
\psi_q (x',y') = g_{h_q} (x',y') \ast h^{-1}_q (x',y').
\end{equation}
The hologram $g_{h_q} (x,y)$ appears in the form of the individual Fresnelets.
Since the continuous subfunction in Eq. (4) is spread over the whole area of the hologram, its aperture will have a lateral full size $L$,
and thus, $g_h (x,y) = \rm{rect} \it (x/L,y/L) h(x,y)$. 
These states are not orthogonal to each other because the overlapping integral between states exists, which represents a quantum-mechanically mixed state.

Considering the rectangular aperture, the state mode has the form of a sinc function distributed with $(\lambda z/\it \Delta x, \lambda z/\Delta y)$ intervals:
\begin{equation}
\psi (x',y') = \mathcal{N} \frac{\sin (\pi L x'/\lambda z)}{\pi L x' / \lambda z} \frac{\sin (\pi L y'/\lambda z)}{\pi L y' / \lambda z},
\end{equation}
where $\mathcal{N}$ is the normalization constant.
The width of the first maximum peak of a sinc function defines an image resolution limit.
The multiple images have the same resolution.

The wave field incident on the hologram acts as individual photons with a momentum, $p=\hbar k$, 
where $\hbar$ is the reduced Planck constant.
The particular spatial mode will be selected by the interaction of photon field with respective Fresnel zone, 
which is a measurement process quantum mechanically.
The spatial resolution of image could be interpreted from the Heisenberg uncertainty relation \cite{24}.
Putting the photon momentum in the lateral direction as $p_{x'} = p \rm{sin}(\it{\Omega}/\rm{2})$, 
the general uncertainty relation is written by
\begin{equation}
\it{\Delta}x' \it{\Delta} p_{x'} \geq \frac{\hbar}{\rm{2}},
\end{equation}
where $p_{x'}$ could be regarded as the maximum uncertainty of photon momentum in the lateral coordinate.
From this,
we find that the spatial resolution of the corresponding images to the replication Fresnel lens has the same value generated from the hologram numerical aperture with a whole aperture window:
\begin{equation}
\Delta x' \cong \frac{h}{2p \sin (\it{\Omega}/\rm{2})} = \frac{\lambda}{\textrm{2NA}}.
\end{equation}
The term on the right side in Eq. (15) was readjusted in a view of the Abbe-Rayleigh diffraction limit constraining spatial resolution for a diffraction-limited imaging system.
It is interesting that using the relation of Eq. (14), the image resolution could be improved beyond classical limit.
Quantum imaging by using correlations between multiple photons has been known to surpass the classical limit \cite{25,26,27}.

Considering a real object, we obtain the following convolution relation for the restored image with the sinc function related to aperture size \cite{8,18}:
\begin{equation}
\sum_{q} o \left(x' + \frac{\lambda zq}{\it{\Delta} x} \right) \ast \rm{sinc} \it{} \left( \frac{\pi L x'}{\lambda z} \right),
\end{equation}
where the prefactors are omitted for convenience.
The respective images are convolved with the Fraunhofer diffraction pattern of hologram aperture window,
where we know that the spatial resolution is defined by the maximum peak of sinc function.
Referring to our previous research for a close relation between the viewing angle and resolution of reconstructed image \cite{8},
we conclude that all reconstructed replica images have the same viewing angle $\it{\Omega}$ of a holographic image in Eq. (1).

The self-similar property of kernel function appearing in the form of modulating curve makes it possible to form well-behaved replica fringes.
The replicated subfunctions are continuous over the entire area of digital hologram.
The incident photon fields on the digital hologram experience a whole area as one aperture.
The optically reconstructed images from the enhanced-NA hologram show the increase of viewing angle,
but there appear the image reduction and the interference of adjacent images when viewing one with a large depth.

\section{Optical hologram imaging for enhanced-NA Fresnel hologram}

Figure 4 illustrates a schematic diagram of the in-line holographic system used to synthesize the enhanced-NA Fresnel hologram at a distance $z_{en}$ lower than $z_c$.
In order to avoid the aliased error in the process of hologram synthesis, the object space should be confined to the diffraction region by a hologram pixel,
and thus, it decreases with decreasing the synthesis distance.

The synthesized digital holograms are shown in Fig. 5.
For convenience, in order to discriminate apparently the high-order fringes, the digital hologram with 256$\times$256 pixels of a pixel pitch 8 $\mu$m is displayed.
We used two letter objects separated from each other in the axial direction and vertically stacked on the coaxial $x$-axis.
This configuration is very useful to measure a smaller viewing angle. 
The viewing angle is measured by observing the maximum parallax.
The Gerchberg-Saxton iterative algorithm is applied to extract the phase hologram \cite{28}.
The four replica fringes in the digital hologram made at half the distance of $z_c$, 15.4 mm are spatially created in Fig. 5(a).
We note that the replica fringe of the former letter is placed at a shifted position vertically or horizontally.
In the reconstruction process, the replica fringes will restore the corresponding images at their respective locations, 
which coincide with the high-order reconstructed images by the pixel pitch of the digital hologram.

Figure 6 shows the optically reconstructed images for the enhanced-NA digital holograms.
We used a phase spatial light modulator (Holoeye PLUTO) with 1920$\times$1080 pixels and a pixel pitch of 8 $\mu$m.
The blue laser of a 473-nm wavelength is utilized as the source of incident plane wave.
The $z_c$-distance for $x$-direction is calculated to be 259.8 mm.
The restored image from the hologram made using the objects located at half the distance of $z_c$, 129.9 mm, is displayed in Fig. 6(a).
Here, two letter objects are located at a distance of 20 mm. Several high-order images are captured within the lens aperture of the camera.
The high-order images of the reconstructed image are placed at the positions specified in the synthesized hologram.
We also observe that the first-order image of the former letter is placed at a horizontally shifted position owing to a different perspective view.
Since the viewing direction of captured image is set to be the zeroth-order image, the images adjacent to the central image show their perspective views.
We can confirm that the perspective view of the central image is changed by moving the viewing direction to the adjacent image, in Fig. 6(b).
The viewing angle is estimated to be 7.1$^\circ$ from the maximum perspective view of the reconstructed image,
which is similar to the value 6.8$^\circ$, calculated from the whole aperture of the hologram.
This means that all images reconstructed optically from the hologram fringes have the same viewing angle with respect to the whole aperture of the digital hologram.
That is, the corresponding aperture size is not limited to the area within the boundary of each fringe.

Figure 6(c) is the restored image for the objects located at a quarter of the distance of $z_c$, 64.9 mm.
For convenience, two letter objects are separated by a distance of 10 mm. 
We can see up to the second-order restored image within the diffraction zone with respect to an object pixel pitch of 2 $\mu$m.
The former letter image is shifted twice in comparison to the first-order image, and the viewing angle of the second images is estimated to be 14.8$^\circ$.
In the restored image from the holograms synthesized using the objects placed at a distance of 32.5 mm, we can see up to the fourth-order images.
Here, the estimated maximum viewing angle is appeared to be 28.1$^\circ$ from two object images separated by 5 mm.
The perspective views of all orders of images are clearly confirmed by changing the viewing direction to the respective images.
Figure 6(d) is a plot of the variations in the viewing angle of holographic image as a function of the reconstruction distance, 
where the simulated curve is based on Eq. (1).
It is noted that the holographic image with a high viewing angle can be obtained at a shorter distance.

\section{Wide viewing-angle holographic display}
\subsection{Method for enlarging viewing-angle of holographic image without sacrificing image size}

For realizing the wide viewing-angle holographic display,
the increase of viewing angle should be feasible without sacrificing the image size.
Furthermore, the interference of high-order images should be prevented.
The sampling condition of a digitized space in the Fresnel hologram synthesis is defined by the Nyquist sampling theorem.
The pixel specifications of both hologram and object planes are as follows,
\begin{equation}
\it{\Delta} x'= \frac{\lambda z}{N \Delta x}.
\end{equation}
In the enhanced-NA Fresnel hologram, although the sampling criterion with respect to the object plane is well satisfied,
the aliased error in the hologram plane occurs.
The point object generates the replica Fresnel zone patterns from the undersampling of digital hologram, as described in Section 2.
However, as studied in our previous work \cite{12},
the replica fringes are suppressed as the point objects are aggregated, and thus, the object with a finite size does not form the replica fringes.
This unusual behavior is resulted from the intrinsic property of the near-field diffraction in the enhanced-NA Fresnel hologram:
\begin{equation}
h(x,y) \textbf{\textit{FT}}\left[o(x',y')h(x',y') \right].
\end{equation}
Above equation is the expanded form about the Fresnel diffraction formula.
The replica fringes are generated from the undersampling of the point spread function $h(x,y)$ in the front of Fourier transform term.
However, the sampling criterion with respect to the object plane is well satisfied.
Namely, the diffraction fringe propagating from the object pixel covers the whole area of digital hologram, which induces the suppression of the replica fringes.

In this case, the digital Fresnel hologram with no aliased fringes could be synthesized even using the extended object out of diffraction region by a hologram pixel, as illustrated in Fig. 7.
Likewise, this feature is inversely applicable to the image reconstruction process,
because the reconstruction process is carried out by the near-field diffraction of the enhanced-NA Fresnel hologram.
The original extended object could be well reconstructed only from the upsampled digital hologram without the interference of high-order images \cite{8}.

Figure 7 presents the configuration of digital hologram synthesis by using the extended object in the enhanced-NA Fresnel diffraction regime.
The distances of $z_1$, $z_2$, and $z_3$ are a half, a quarter, and one eighth of $z_c$, respectively.
The critical distance $z_c$ is 30.8 mm for this specification.
All the blue boxes indicate the diffraction range by a hologram pixel pitch, 8 $\mu$m at respective object spaces.
Real-valued digital hologram computed at a distance, 15.4 mm of a half of $z_c$ is displayed,
where the diffraction scope in the object plane is two times smaller than the hologram extent, based on Eq. (17).
The Lena image with 512$\times$512 pixels of a 4-$\mu$m resolution has a double size of 1024-$\mu$m diffraction scope.
Nevertheless, the computed hologram does not show an aliased fringe due to the near-field diffraction property.
Similarly, no aliased fringes are formed in the digital hologram made at distances of $z_2$ and $z_3$. 
Here, the whole area of digital hologram would act as one aperture despite the suppression of replica fringes. 
The hologram fringe suppression is unlike a spatial filtering occurring during the hologram synthesis by a convolutional or angular spectrum method [8].
The method using a double Fourier transform filters out the diffraction fringe by selecting the zeroth-order spectrum in the intermediate Fourier space.

The digitized Fresnel hologram reconstructs the high-order images at a shifted position of $\lambda z/\it{} \Delta x$ in the object plane,
which could appear in the form of modulating a high frequency wave \cite{18}: 
\begin{equation}
\sum_{q} o \left(x' + \frac{\lambda zq}{\it{\Delta} x} \right) \rm{exp} \it{} \left( i\frac{\rm{2} \it{} \pi qx'}{\Delta x} \right).
\end{equation}
The corresponding frequency of high-order image is $q/\it{} \Delta x$.
Let us represent the carrier frequency by a pixel pitch $\it{} \Delta x'$ of object.
In the conventional Fresnel hologram made at a distance larger than $z_c$, the resolution of object plane is lower compared to the hologram resolution,
where the high frequency term would disappear because the required frequency is greater than the Nyquist frequency of object plane.
Thus, the high-order image does not have a modulating form in a general simulation situation.
However, if the sampling rate of object is finer, the modulating form is regenerated.
Meanwhile, the enhanced-NA hologram reconstructs the object with a finer resolution compared to the hologram resolution.
When the hologram is made at a distance of a half of $z_c$, the pixel interval $\it{} \Delta x'$ is calculated to be $\it{} \Delta x/\rm{2}$, from Eq. (17),
where the spatial frequency of a shifting-frequency wave becomes $1/(\rm{2} \it{} \Delta x')$.
The first-order image is modulated with a high-frequency wave of $1/(\rm{2} \it{} \Delta x')$.
The spatial frequency is determined by a geometrical structure.

Figure 8(a) is the restored image from the digital hologram made at a distance of a half of $z_c$, in Fig. 7.
The restored image is mixed with the modulated high-order noises.
That is because the diffraction scope by a hologram pixel is a half of the image size of 2048 $\mu$m.
Here, we can expect that if the hologram pixel is upsampled to be 4 $\mu$m, the high-order noises are effectively removed.
As depicted in Fig. 7, the upsampling process enlarges two times the diffraction scope, indicated by a diffraction angle $\theta_{\rm{u1}}$.
Since the diffracted field from the upsampled hologram pixel occupies a full extent of image plane,
the first-order image signals become a kind of the replica noises within a diffraction scope,
which would be suppressed due to the near-field diffraction property.
Figure 8(b) shows that the twofold upsampled hologram well removes the first-order noises to obtain the original Lena image.

The intensity and real value profiles in the horizontal direction at the center of image are drawn in Figs. 8(e) and 8(f), respectively.
It looks like that the sinusoidal wave is clearly filtered out by an upsampling process.
But, this process is fundamentally different from the simple spatial filtering.
As stated previously, although the high frequency term is deleted by upsampling a hologram pixel, the increase of object resolution could prohibit this removal.
Optically, the object space is continuous.
%Furthermore, the high-order terms survive despite a removal of high frequency wave, in Eq. (19).
We confirmed that the suppression of high-order images arises irrespective of the object resolution even in a numerical simulation.
For this reason, we conclude that the property of the near-field diffraction of the enhanced-NA Fresnel hologram allows it to implement this extraordinary phenomenon.

Figure 8(c) shows the well restored images from the digital hologram computed at a distance $z_2$, 7.7 mm,
where the object size is four times larger than the diffraction scope.
The digital hologram is upsampled fourfold to diffract the incident wave to object size.
Restored image from the digital hologram made at a distance $z_3$ is displayed in Fig. 8(d).
The eightfold upsampled hologram clearly reconstructs the Lena image,
where its viewing-angle gets to 30.9$^\circ$, from Eq. (1).
We find that it is possible to restore an original image even at a finer upsampling process for a high-NA hologram without higher-order noises.
The upsampling process in optical experiment can be performed by using a simple grid.
We surely know that the high-order image reappears outside the diffraction range by an upsampled hologram pixel.
The finer upsampling would lead to a larger diffraction zone preventing high-order noises.
Figure 9 shows the restored images from the synthesized hologram at a distance $z_1$ in the extended image space.
The replica images are distributed at intervals of the diffraction range by a twofold upsampled hologram pixel,
which is effectively suppressed by a fourfold upsampling process.
As studied previously [8], above result supports that a wide viewing-angle holographic display could be realized even by using a commercial spatial light modulator.

\subsection{Optical experiments for verification}

To conduct the upsampling process without attaching a finer grid to the pixelated modulator,
we utilize a low-resolution digital hologram as compared to a pixel resolution of modulator.
Most of the specifications of optical experiments are set to be those for hologram imaging in Section 3.
The digital hologram is prepared through two separate steps.
Firstly, the hologram with 1920$\times$1080 pixels and 8-$\mu$m pixel pitch is calculated by using the object placed at a distance of 259.8 mm,
where both the object and the hologram have the same pixel specifications.
The adequate phase hologram is generated by using the Gerchberg-Saxton iterative algorithm.
Secondly, the digital hologram is undersampled by twice the pixel interval, and finally,
it consists of 960$\times$540 pixels with a pixel pitch of 16 $\mu$m.
The digital hologram loaded into the spatial light modulator with 1920$\times$1080 pixels is obtained by a duplication of adjacient pixel value.
In the pixelated modulator, the inner pixel boundaries between two pixels having the same value play a role in an upsampling grid.

Figure 10 is the calculated digital hologram and the reconstructed image showing an extended field of view.
The synthesis distance of 259.8 mm is a half of critical distance for this specification, in Eq. (17).
The diffraction angle by a hologram pixel pitch of 16 $\mu$m is 1.7$^\circ$.
The object space is two times larger than the lateral diffraction range.
As shown in Fig. 10(a), the replica fringes are not formed despite using the iterative algorithm
when the object space is fully occupied by a object content without a marginal vacant space. 
If the upsampling grid does not exist, the high-order noises interfere with a proper image as like Fig. 8(a).
However, as illustrated in Fig. 10(b), we can view the holographic image without the interference of high-order noises
at the viewing angle of 3.4$^\circ$, which is twice the diffraction angle by a hologram pixel pitch.

\section{Discussions}

Optical reconstruction strongly depends on the performance of spatial light modulator.
The present spatial light modulators are capable of only the phase or amplitude modulation.
An amplitude or complex modulation could be applicable to this type of wide viewing-angle holographic display system.
Complex hologram was used in this numerical simulation, and thus, it would be desirable to execute a complex modulation \cite{29,30}.

In pratice, the phase hologram has attracted attention owing to a high diffraction efficiency.
The practical phase hologram could be obtained in terms of the iterative optimization algorithms.
Furthermore, the object content is placed in the vacant space in general.
In such cases, since the replica fringes are formed during synthesizing an enhanced-NA hologram,
the generation of high-order images is unavoidable, as described in Section 3.
The replica fringe is formed due to the sampling rate lower than the proper sampling rate.
We know that this phenomenon arises from the sub-sampling by a multiple pixel pitch, as displayed in Fig. 1.
If the sub-sampling is nonuniformly performed, there appear no replica fringes.
If there is also an irregularly pixelated spatial light modulator which has nonuniformly distributed pixels of irregular pixel size,
it would effectively hinder the generation of high-order images.
Another simple method is to use the spatial filter having nonuniformly distributed pixel aperture, attached to the pixelated modulator.
These also make it possible to realize the wide viewing-angle holographic display, but the optimization procedure for the hologram
synthesis is needed because of an irregularly pixelated hologram.
Recently, various studies have been performed to synthesize the phase hologram without the iterative algorithm or random-phase addition \cite{31}.

From a different point of view, we find that on the basis of our interpretation for viewing angle of holographic image, a finer resolution of image would lead to a wider viewing-angle. 
In a classical wave optics, the typical methods to increase the image resolution are to reduce the wavelength or to build higher numerical-aperture optics. 
However, in a quantum imaging system, the Abbe-Rayleigh diffraction limit is secondary to the problem at hand, that is to say,
it can achieve the ultimate optical resolution \cite{25,26,27}.
Quantum imaging by using $N$-correlated photons leads to a $1/\sqrt{N}$ resolution enhancement called standard quantum limit.
Furthermore, for $N$ entangled photons one can achieve a $1/N$ scaling known as the Heisenberg limit.

If the image with higher resolution beyond the Rayleigh limit can be floated in a free space, the viewing angle would be larger than the angle estimated from the system NA.
The image spot can be regarded as only a point source generating secondary wavelets, 
and thus, the diffraction angle would follow from the diffraction formula, $\it{\Omega} =\rm \sin^{-1} (\it \lambda/ \Delta x')$. 
The smaller image spot generates a wave with a higher angular spectrum. 
We find that the image resolution becomes a primary measure in defining the viewing angle of holographic image. 
The Fourier hologram generates the image with a constant spatial resolution dependent on only the focal length of lens, irrespective of image depth \cite{32}. 
The digital hologram synthesized by a convolutional or angular spectrum method reveals a constant image resolution at a distance longer than a critical distance \cite{8}. 
These properties of spatial resolution are not directly explained by a classical system NA.
Here, we could infer the effective NA from the image resolution to interpret consistently this phenomenon.
For a practical application based on quantum imaging,
the effective imaging techniques without complicate processes such as a coincidence or multi-photon measurement should be developed.

\section{Conclusions}

The viewing angle of holographic image in the enhanced-NA Fresnel hologram is well understood from the analysis of the internal structure of the sampled PSF.
The angle value is determined by the NA of whole aperture of digital hologram.
Optical experiment shows the consistent result with the interpretation of viewing angle dependent on the hologram numerical aperture.
The high-NA hologram reconstructs the image at a higher viewing-angle, while the image reduction is inevitable due to the Nyquist sampling criterion.
The enhanced-NA Fresnel diffraction has an interesting characteristic that the aliased replica noises are effectively suppressed within the diffraction scope by a digitized pixel.
The digital Fresnel hologram synthesized even using the extended object out of diffraction region shows no aliased fringes.
Only the upsampling process of digital hologram results in the image reconstruction without the interference of high-order images.
This approach gives the method for enlarging the viewing angle of holographic image without sacrificing image size.

This work was partially supported by Institute of Information \& Communications Technology Planning \& Evaluation (IITP) grants funded 
by the Korea government (MSIP) (2017-0-00049 and 2021-0-00745)

\begin{figure}
\includegraphics[scale=0.85, trim= 1cm 7.5cm 1cm 0cm]{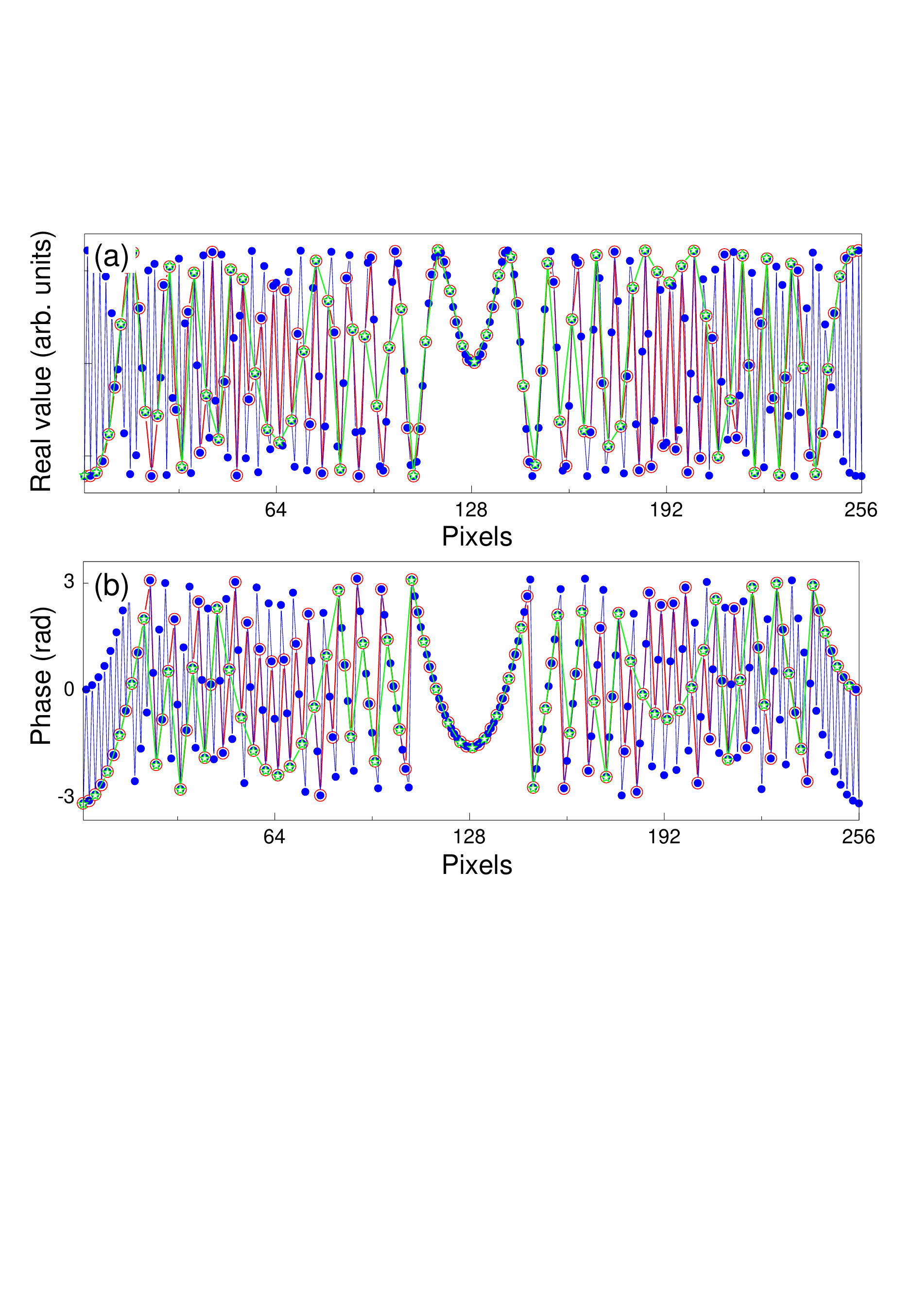}
\caption{Quadratic phase curves of (a) real-valued and (b) angle-valued PSFs are drawn in one-dimensional digitized space.
The speciﬁcations of the blue curve are as follows; wavelength $\lambda$ = 532 nm, distance $z = 30.8$ mm, pixel number $N = 256$, and pixel pitch $\it{\Delta} x = \rm{8}$ $\mu$m.
The red and green curves are sampled by pixels of $2\it{\Delta} x$ and $4\it{\Delta} x$, respectively.}
\end{figure}

\begin{figure}
\includegraphics[scale=0.8, trim= 1cm 10cm 0.0cm 1cm]{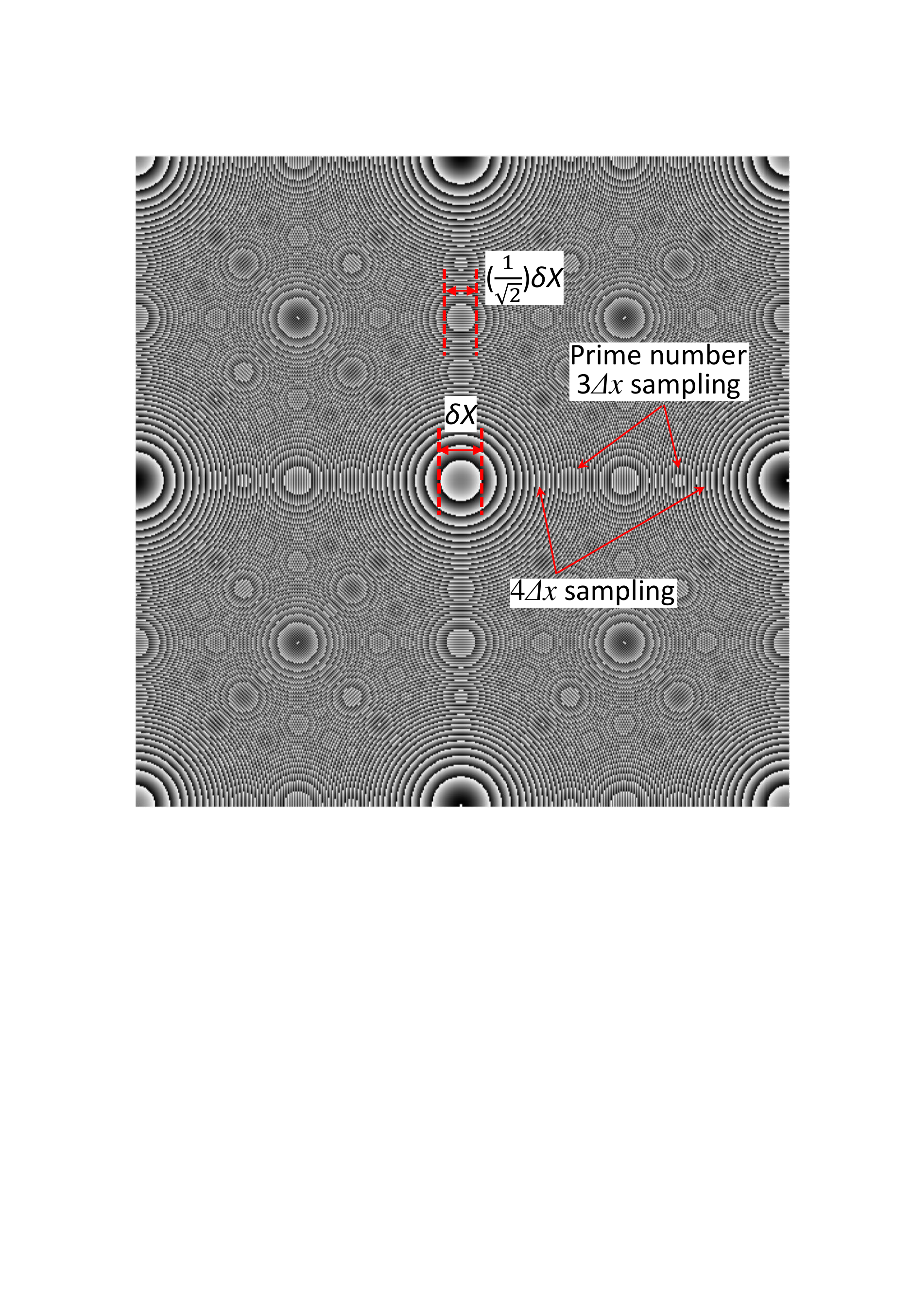}
\caption{Fractal structure of the sampled two-dimensional PSF.
Angle-valued two-dimensional PSF, known as Fresnel zone plate, are drawn at specifications; distance of 7.7 mm and 512$\times$512 pixels with a pixel pitch of 4 $\mu$m.
Four aliased fringes are generated in this specification, and $\delta X$ indicates the width of core shell of the primary Fresnelet.}
\end{figure}

\begin{figure}
\includegraphics[scale=0.85, trim= 2.5cm 4.5cm 3cm 1cm]{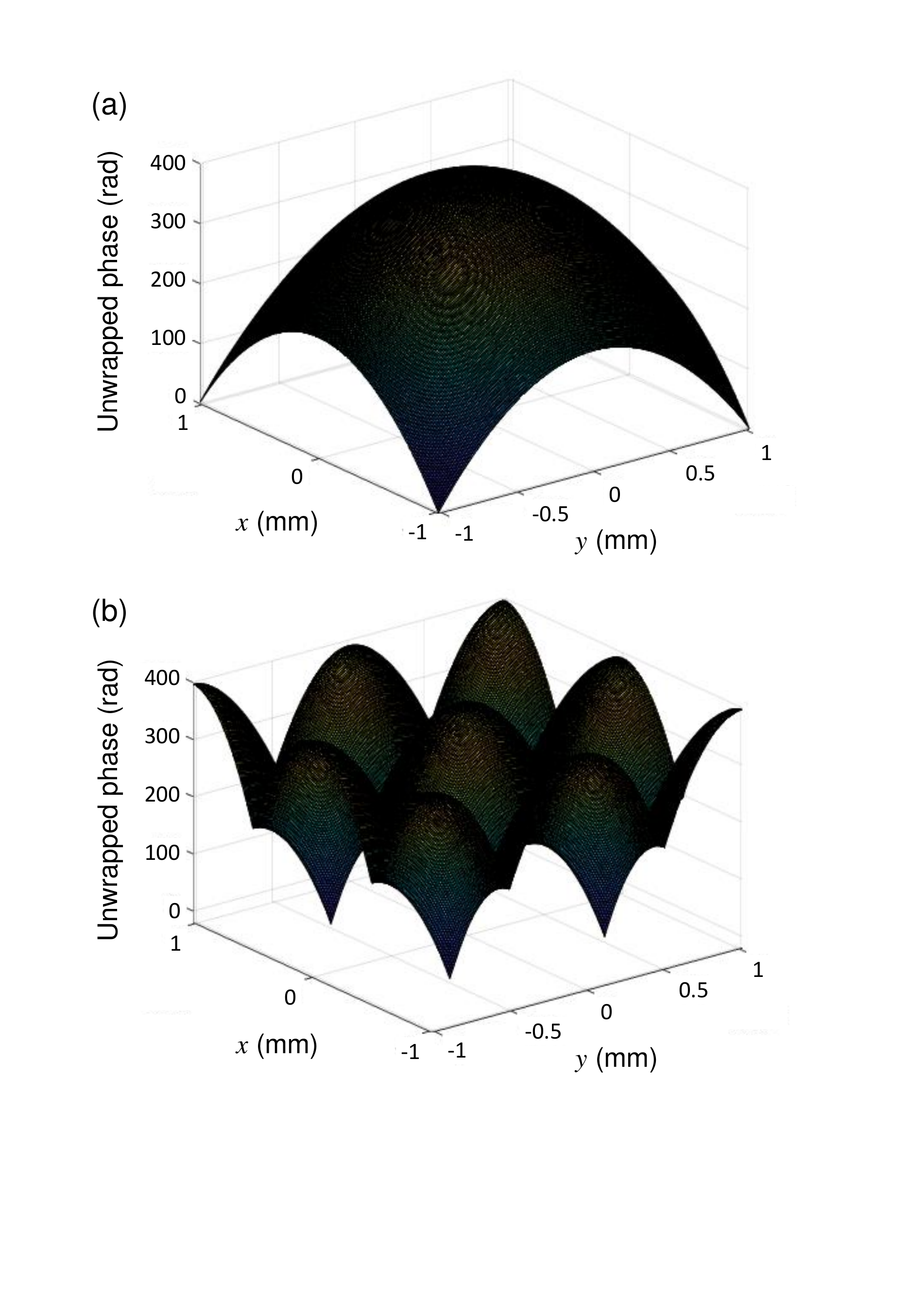}
\caption{Unwrapped phase of Fresnel zones synthesized at (a) a critical distance $z_c$ and (b) a half of distance $z_c$.
The speciﬁcations are as follows; wavelength $\lambda$ = 532 nm, distance $z_c$ = 30.8 mm, pixel number $N$ = 256, and pixel pitch $\it{\Delta} x = \rm{8}$ $\mu$m.}
\end{figure}

\begin{figure}
\includegraphics[scale=0.95, trim= 0.5cm 11cm 1cm 2cm]{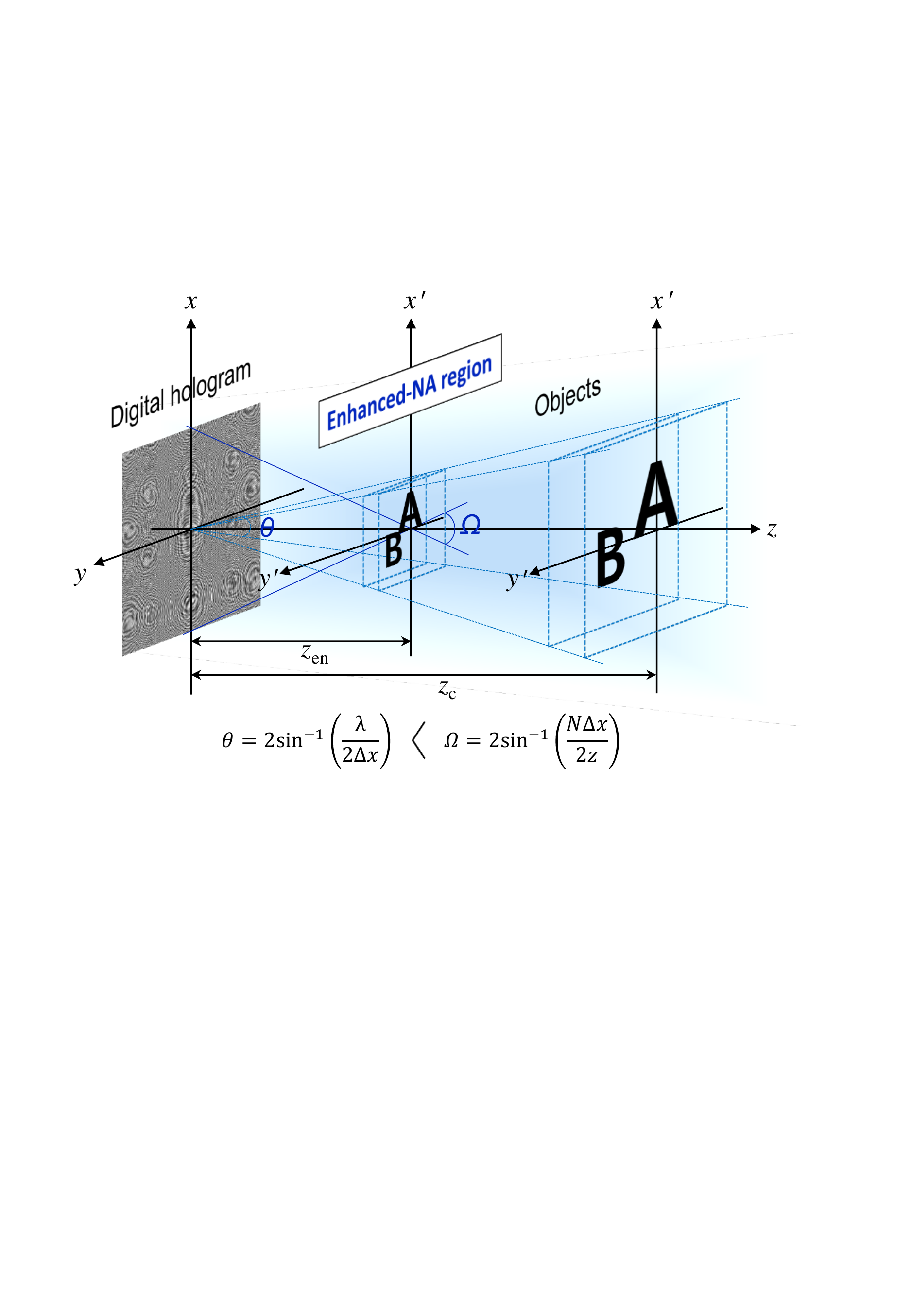}
\caption{Schematic of enhanced-NA Fresnel hologram synthesis via the in-line holographic system.
The physical sizes of the hologram and object are the same at a distance $z_c$.
The amplitude hologram synthesized by Gerchberg-Saxton iterative algorithm is displayed.
$\theta$ is the diffraction angle by a pixel pitch $\it{\Delta} x$ of digital hologram.}
\end{figure}

\begin{figure}
\includegraphics[scale=0.85, trim= 1cm 14cm 1cm 2cm]{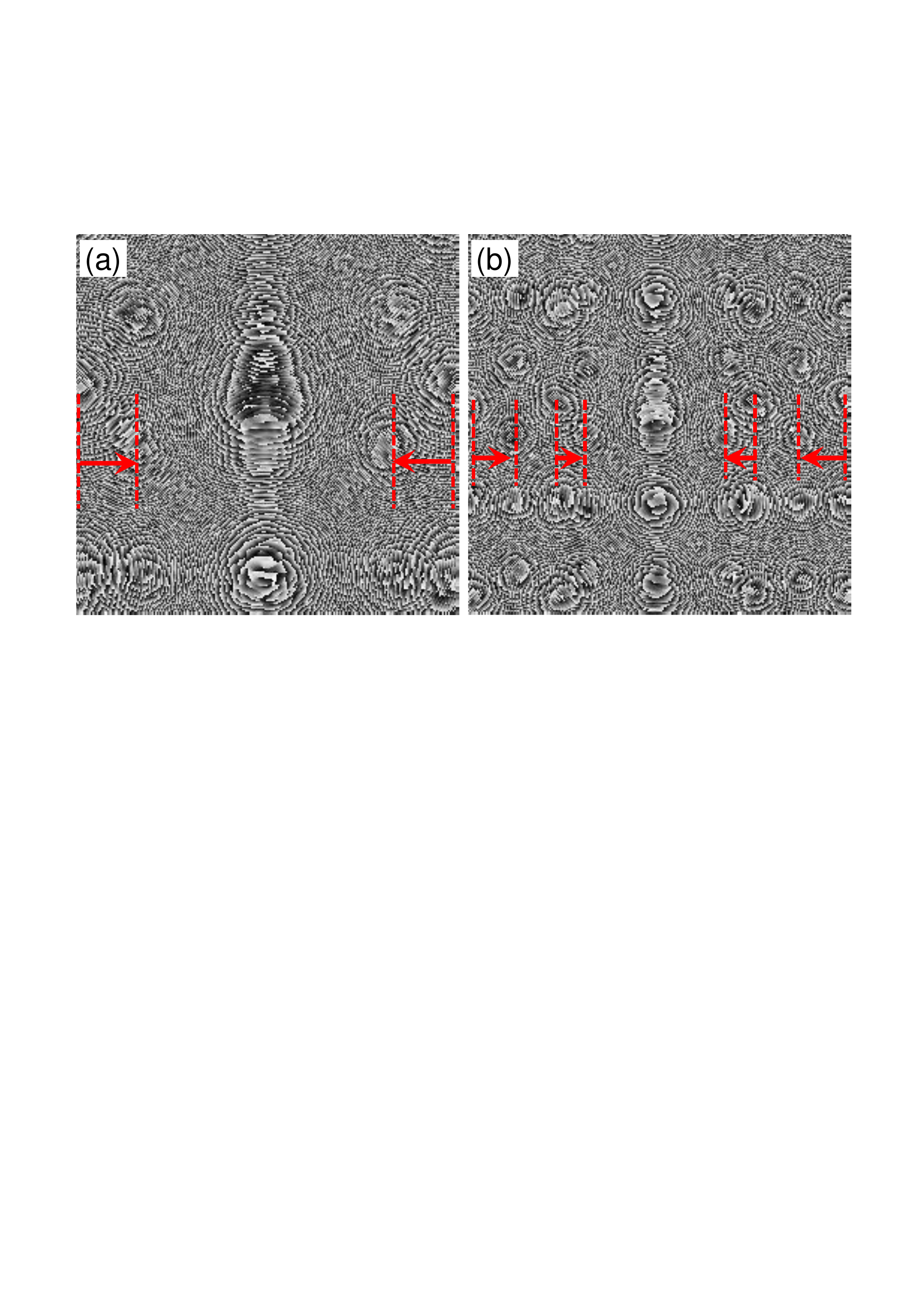}
\caption{Enhanced-NA Fresnel hologram synthesized through the in-line holographic system.
The critical distance $z_c$ is 30.8 mm for this specification.
(a) Phase hologram of two letter objects synthesized at half the distance of $z_c$ by the Gerchberg-Saxton iterative algorithm with 30 steps.
Two letters are separated by 5 mm.
The red arrow indicates the shift of the ﬁrst letter fringes. 
(b) Phase hologram synthesized at a quarter of the distance of $z_c$. Two letters are separated by 2 mm.}
\end{figure}

\begin{figure}
\includegraphics[scale=1, trim= 1.2cm 8cm 1cm 2cm]{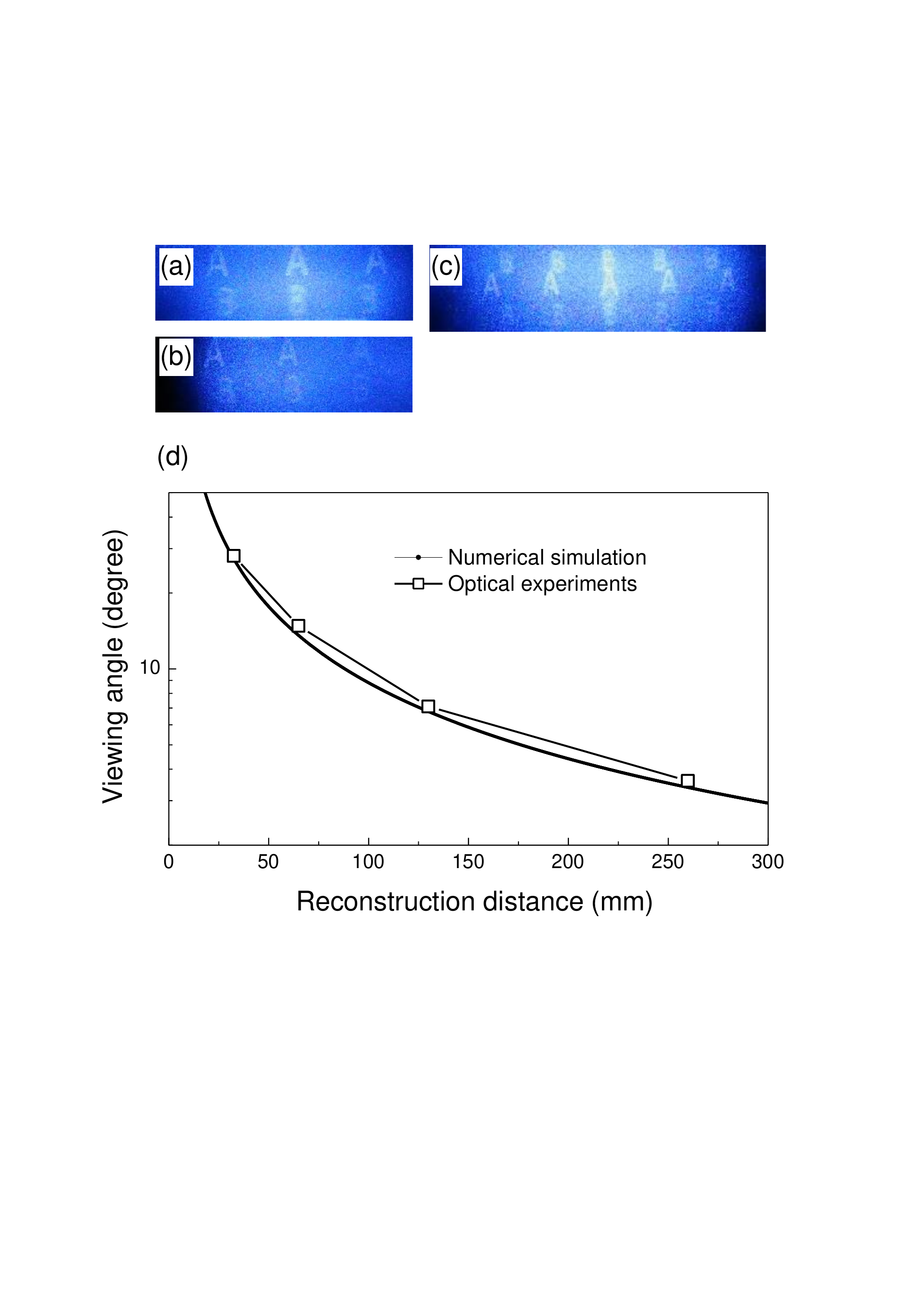}
\caption{Optically reconstructed images and their viewing angle variation for the enhanced-NA Fresnel hologram.
The images are captured at a slightly inclined vertical-direction to avoid the directed beam.
(a) Reconstructed image of the digital hologram made at a distance of 129.9 mm.
(b) Reconstructed image captured in the viewing direction of the first-order image.
(c) Reconstructed image for the digital hologram made at a distance of 64.9 mm.
(d) Viewing-angle variations of the holographic image as a function of the reconstruction distance.}
\end{figure}

\begin{figure}
\includegraphics[scale=1, trim= 1.5cm 12cm 1cm 2cm]{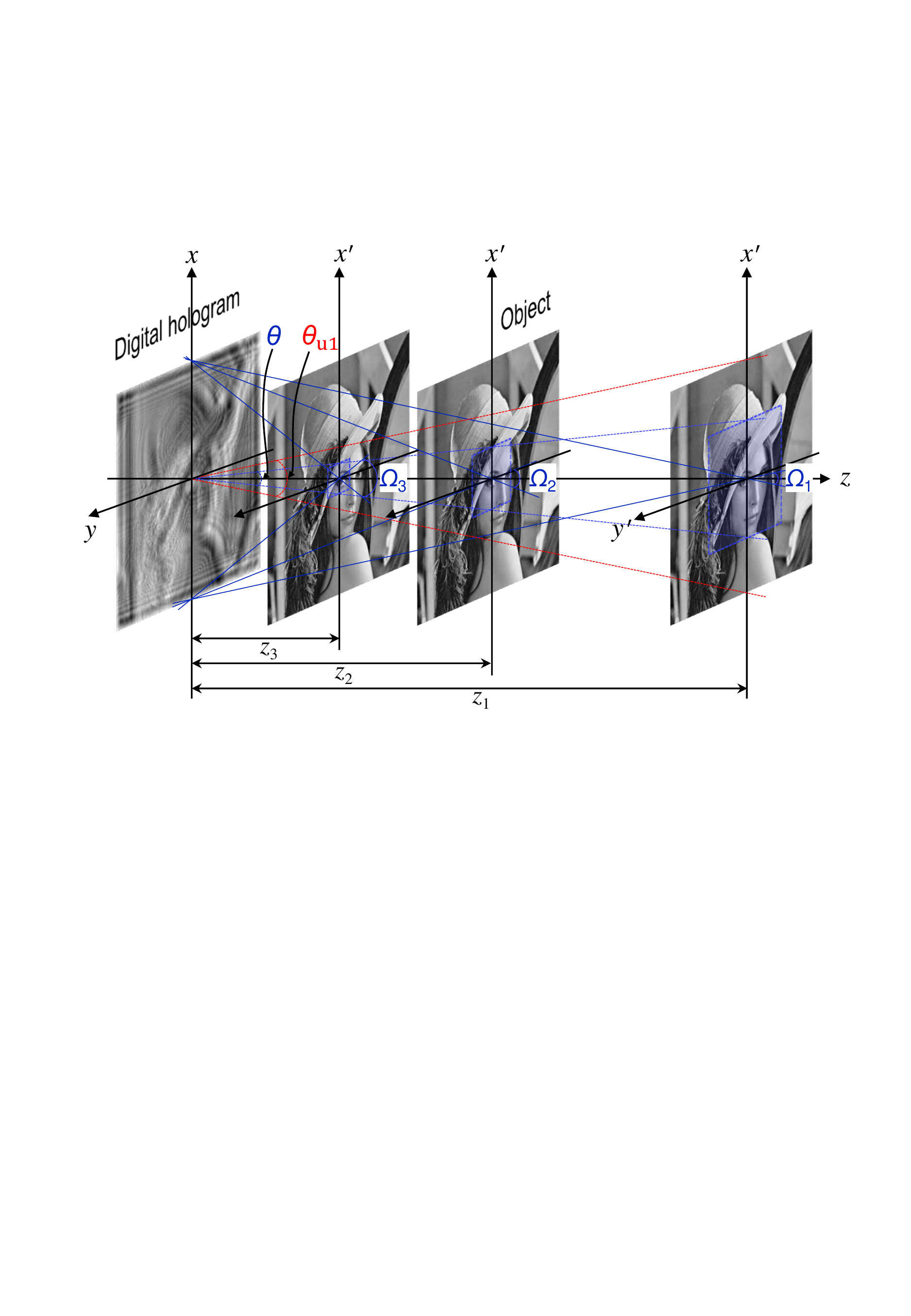}
\caption{Configuration of the enhanced-NA Fresnel hologram synthesis by using an extended object out of a diffraction scope.
The speciﬁcations of hologram synthesis are as follows; wavelength $\lambda$ = 532 nm, pixel number $N$ = 256, and hologram pixel 8 $\mu$m.
The distances of $z_1$, $z_2$, and $z_3$ are a half, a quarter, and one eighth of $z_c$ of 30.8 mm, respectively. 
Real-valued digital hologram computed at a distance of a half of $z_c$, 15.4 mm is displayed.
The diffraction angle $\theta$ by a hologram pixel is calculated to be 3.8$^\circ$,
where all the blue boxes indicate the diffraction range by a hologram pixel pitch at respective object spaces.
The red line relates to the diffraction angle $\theta_{\rm{u1}}$ of 7.6$^\circ$ by upsampling pixel to be 4 $\mu$m.
The viewing angles of $\it{\Omega}_{\rm{1}}$, $\it{\Omega}_{\rm{2}}$, and $\it{\Omega}_{\rm{3}}$ are calculated to be 7.6$^\circ$, 15.3$^\circ$, and 30.9$^\circ$, respectively.}
\end{figure}

\begin{figure}
\includegraphics[scale=0.95, trim= 1cm 6.5cm 1cm 3cm]{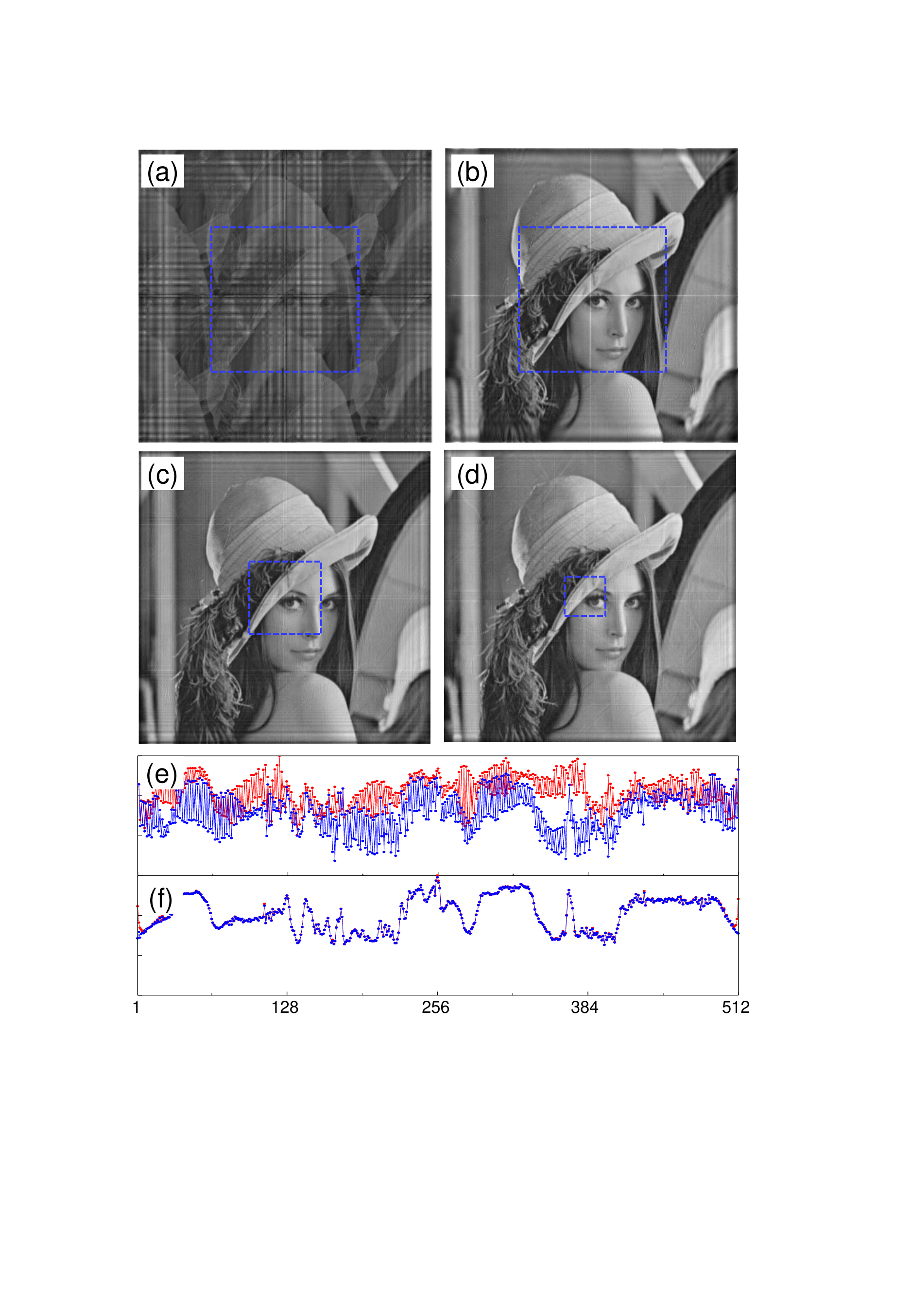}
\caption{Characteristics of reconstructed images from the enhanced-NA Fresnel hologram synthesized by using an extended object.
All blue boxes indicate the diffraction range by a hologram pixel, $N \Delta x' = \lambda z/\it{} \Delta x$.
Restored images from the digital hologram made at a distance $z_1$ (a) without an additional upsampling process and (b) with two-fold upsampling process.
(c) Restored image from the fourfold upsampled hologram made at a distance $z_2$. 
(d) Restored image from the eightfold upsampled hologram made at a distance $z_3$. 
Red and blue lines of graphs are the intensity and real value profiles in the horizontal direction at the center of image.
(e) The upper and (f) lower graphs correspond to (a) and (b) images, respectively}
\end{figure}

\begin{figure}
\includegraphics[scale=1, trim= 2cm 12cm 1cm 3cm]{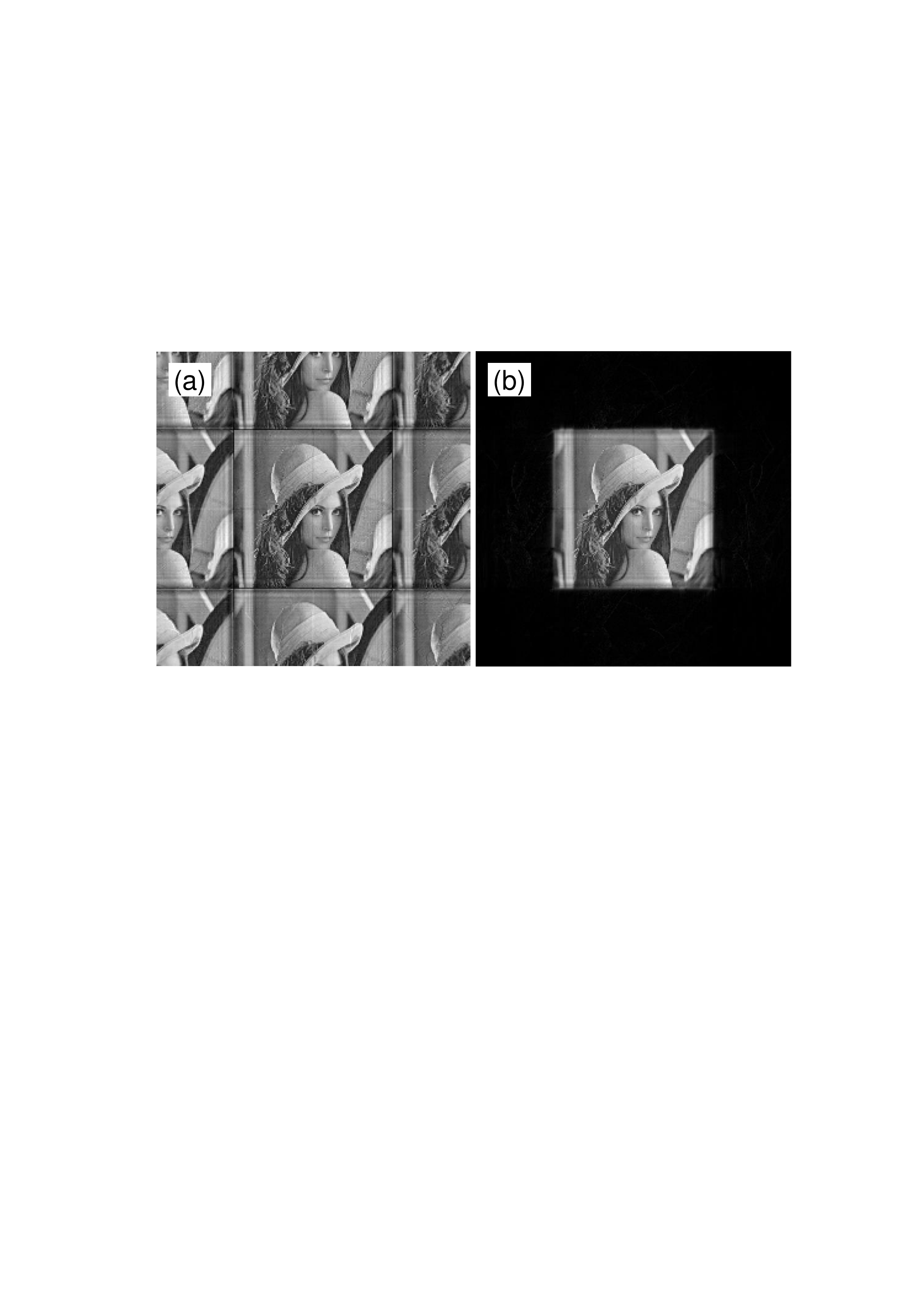}
\caption{Reconstructed images in the extended image space.
(a) Replica Lena images are distributed at intervals of the diffraction range by a twofold-upsampled hologram pixel. 
(b) The fourfold upsampling secures a larger diffraction zone preventing high-order noises.}
\end{figure}

\begin{figure}
\includegraphics[scale=1, trim= 2cm 16.5cm 1cm 0cm]{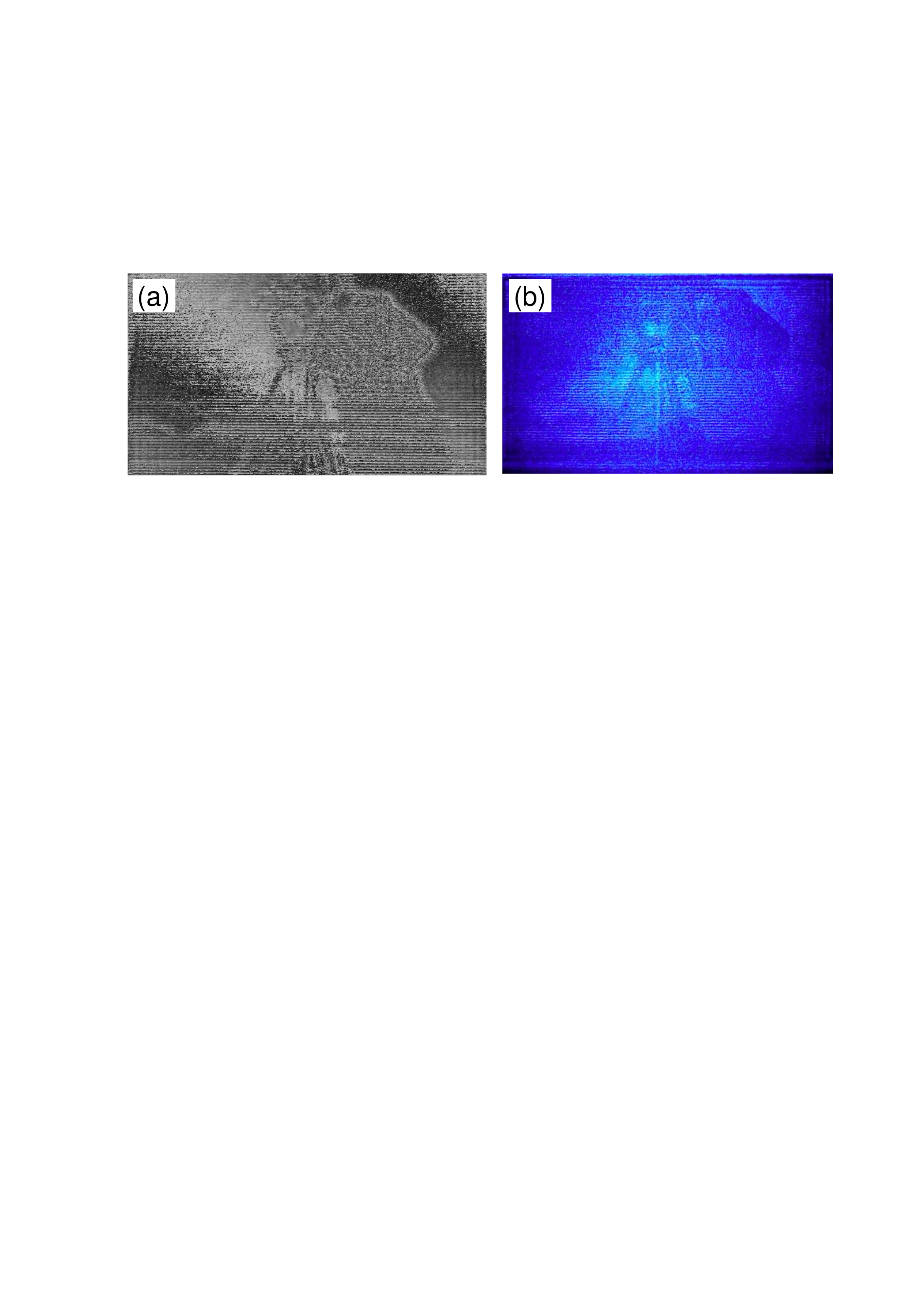}
\caption{Optical experiments for enlarging the viewing angle of holographic images without sacrificing image size.
(a) The phase hologram is calculated using camerman image with 8-$\mu$m pixel pitch.
(b) The reconstructed image without an interference of high-order noises is captured.}
\end{figure}

\end{document}